\documentclass{ws-mpla}
\usepackage[super]{cite}
\usepackage{graphicx}
\begin{document}

\markboth{Authors' Names}
{Instructions for Typing Manuscripts (Paper's Title)}

\catchline{}{}{}{}{}

\title{LATTICE GAUGE THEORY AND GLUON COLOR-CONFINEMENT IN CURVED SPACETIME\\
}

\author{\footnotesize KRISTIAN HAUSER VILLEGAS
}

\address{National Institute of Physics, University of the Philippines, Diliman\\
Quezon City, 1101,
Philippines
\\
kvillegas@nip.upd.edu.ph}

\author{JOSE PERICO ESGUERRA}

\address{Theoretical Physics Group, National Institute of Physics, University of the Philippines Diliman\\
Quezon City, 1101, Philippines
}

\maketitle

\pub{Received (Day Month Year)}{Revised (Day Month Year)}

\begin{abstract}
The lattice gauge theory for curved spacetime is formulated. A discretized action is derived for both gluon and quark fields which reduces to the generally covariant form in the continuum limit. Using the Wilson action, it is shown analytically that for a general curved spacetime background, two propagating gluons are always color-confined. The fermion-doubling problem is discussed in the specific case of Friedman-Robertson-Walker metric. Lastly, we discussed possible future numerical implementation of lattice QCD in curved spacetime.

\end{abstract}

\ccode{11.15.Ha: 04.62.+v}

\section{Introduction}
\label{sec:intro}

There are several reasons for studying QCD in curved spacetimes. One is the possibility of the existence of dense quark matter at the core of a neutron star\cite{Pines}. Although perturbative QCD works well at very high densities (large quark chemical potential $\mu$) where the quark matter is predicted to be in color-flavor-locked phase(CFL)\cite{Alford}, we don't have the same confidence in the regime intermediate  between the hadronic phase and the CFL phase (called non-CFL phase). In this regime, we expect a strongly coupled QCD. Further, due to the extreme density involved, the effects of spacetime curvature can not be neglected. Hence, a proper treatment of this regime needs the full treatment of non-perturbative QCD method in curved spacetime. This is the major motivation of this work. We want a lattice QCD that is valid for curved spacetimes.

This paper also partially addresses the question of the possibility of QCD deconfinement in a curved spacetime. There are several reasons for considering this question. 1.) In curved spacetime formulation of QFT, the metric enters non-trivially in the action: contraction of vector indices, Christoffel and spin connections and in covariant derivatives of vector bosons. Because of this, one might ask: will the linear-in-separation potential between two colors in flat spacetime get modified in curved spacetime? Is it possible to find a metric $g_{\mu\nu}(x)$ that will significantly modify this potential so that it will lead to deconfinement? 2.) As pointed-out by \cite{Gass}, in flat spacetime, QCD undergoes a confining-deconfining phase transition at high temperatures. However spacetime curvature can play the role of temperature\cite{Davies81}. In blackholes, for example, the temperature is inversely proportional to the surface gravity \cite{Hartle}. In fact, using the quark condensate as the order parameter in Nambu-Jona-Lasinio model, it was shown that there can be deconfinement just outside the blackhole horizon.\cite{Flachi13} 3.) Spacetime curvature can lead to particle creation. For large curvatures, the quark-gluon plasma created might screen the color flux leading to deconfinement.

From these considerations, it is therefore interesting to study the effects of gravity on QCD confinement.

Due to the complexity of the problem, investigations usually proceed by using a simpler model. For example, \cite{Gass} uses the Schwinger model in mock Schwarzschild metric in 1+1 dimensional spacetime. However, this model did not resolve the problem. As the author himself pointed out, the model has many shortcomings: 1.) the Schwinger model in flat spacetime is confining even at high temperatures 2.) it uses an abelian gauge field and 3.) it is superrenormalizable.

Some papers\cite{Gaete, Gaete06, Gaete07} use a phenomenological model wherein a dilaton is coupled to gluodynamics to investigate the problem and they were able to derive the Cornell potential\cite{Eichten} between the quark-antiquark pair.  However, these papers restricted their treatment to the weak coupling limit. Also, the coupling to dilaton means that their approach focused on the effect of quantum gravity instead of the classical curved spacetime background to confinement, which is the concern of the current paper. 

Vasihoun et. al. investigated the effects of gravity and topology on non-linear gauge field theory.\cite{Vasihoun14} They have shown that there can be charge \lq\lq hiding\rq\rq in a wormhole throat. When coupled to a dilaton, they found confinement-deconfinement effect as a functions of dilaton VEV.

We must also mention that in the literature, there are other motivations for studying QCD in curved spacetime. For example, it can be used as a mathematical trick to track the symmetries by introducing a new field\cite{Kharzeev04, Migdal}. In this approach, the gluons are coupled to the dilaton, and the following duality was obtained: classical gluodynamics in curved conformal spacetime backgrounds $\Leftrightarrow$ gluodynamics in flat spaetime coupled to scalar glueballs.

Sasagawa and Tanaka are concerned with the direct effect of gravity on QCD\cite{Sasagawa}. They investigated the chiral and deconfinement phase transitions in curved spacetimes with $R\times S^3$ and $R\times H^3$ topology. This has been done by using the Nambu-Jona-Lasinio model with the dressed Polyakov loop as the order parameter for the deconfinement phase transition. Their paper is concerned with the effect of gravity on the difference between chiral and deconfinement transition temperatures.

In flat spacetime, lattice gauge theory(LGT) has successfully dealt with color confinement which is a non-perturbative strong-coupling phenomenon. One obstacle to extending LGT to curved-spacetime QCD problems is the need to derive the correct generally covariant form of the action when the continuum limit is taken. In this paper, we will show that a simple generalization of the link and plaquette variables will produce the correct action in the continuum limit. For the quark field, this is done by discretizing the generally-covariant Dirac equation and solving the resulting propagator.

\section{Gluon in a Lattice}
\subsection{Wilson Action}
Just as in the flat spacetime case, we will perform a Wick rotation in the time coordinate $t\rightarrow it$ so that the metric will be Euclidean.\\ 
The link variable can be defined in curved spacetime by inserting the metric in appropriate places
\begin{eqnarray}
U_{\epsilon}(x)=1-i\lambda g_{\mu\nu}(x)\hat{A}^{\mu}(x)\delta x^{\nu}
\end{eqnarray}
where the $\epsilon$ labels the direction of transport, $\lambda$ is the coupling, $g_{\mu\nu}(x)$ is the metric, $\hat{A}^{\mu}(x)$ is the matrix gauge field and $\delta x^{\nu}$ is an infinitesimal displacement.\\
For color transport at finite distance $x$ to $x+a$, the link variable is the path-ordered product of Eq.(2.1)
\begin{eqnarray}
U_{\epsilon}(x)=\mbox{P}\exp\bigg[ -i\lambda\int_{x,\Gamma}^{x+a}g_{\mu\nu}(x)\hat{A}^{\mu}(x)\mbox{d}x^{\nu}\bigg]
\end{eqnarray}
for some specified path $\Gamma$.\\
It can be seen that taking the adjoint produces the reversed-path color transport.

Under the gauge transformation of the gauge field 
\begin{eqnarray}
\hat{A}_{\epsilon}(x)\rightarrow G(x)\hat{A}_{\epsilon}(x)G^{\dagger}(x)-\frac{i}{\lambda}(\partial_{\epsilon}G(x))G^{\dagger}(x).
\end{eqnarray}
the link variable changes as
\begin{eqnarray}
U'_{\epsilon}(x)=G(x+\delta x)U_{\epsilon}(x)G^{\dagger}(x).
\end{eqnarray}
To show this, we used the expansion $G(x+\delta x)\approx G(x)+g_{\mu\nu}\delta x^{\nu}\partial^{\mu}G(x)$ and the fact that $\delta x$ is infinitesimal so that $\mathcal{O}(\delta x^2)$ can be ignored.\\
With this definition of link variables, the plaquette operator can now be defined in an analogous manner with the flat spacetime case

\begin{eqnarray}
P_{\delta x,\delta y}\equiv U_{\delta x}(x)U^{\dagger}_{\delta y}(x)U^{\dagger}_{\delta x}(x+\delta y)U_{\delta y}(x+\delta x).
\end{eqnarray}
The main differences are the presence of the metric and the covariant derivatives when the exponents of the link variables are expanded. For example, one leg of the plaquette has an associated link variable
\begin{eqnarray}
U_{\delta x}(x)&=&e^{-i\lambda g_{\mu\nu}(x+\frac{1}{2}\delta x)\hat{A}^{\mu}(x+\frac{1}{2}\delta x)\delta x^{\nu}}\\
&\approx& e^{-i\lambda g_{\mu\nu}(\hat{A}^{\mu}+\frac{1}{2}g_{\alpha\beta}\delta x^{\alpha}\nabla^{\beta}\hat{A}^{\mu})\delta x^{\nu}}
\end{eqnarray}
where in the last line the metric and the gauge field are functions only of $x$. 

Eq.(2.7) was obtained from Eq.(2.6) by expanding $g_{\mu\nu}(x+\frac{1}{2}\delta x)$ and $\hat{A}^{\mu}(x+\frac{1}{2}\delta x)$ in powers of $\delta x$. Note that $\hat{A}^{\mu}$ carries a contravariant vector index so that in curved spacetime, one must take the full covariant derivative $\nabla_{\alpha}\hat{A}^{\beta}\equiv \partial_{\alpha}\hat{A}^{\beta}+\Gamma^{\beta}_{\alpha \gamma}\hat{A}^{\gamma}$ instead of just a simple partial derivative as is done in Minkowski spacetime. We also used the fact that the covariant derivative of the metric vanishes.

Using the gauge transformation of the link variable Eq.(2.4) and the cyclic property of the trace it can be seen that the trace of the plaquette operator is gauge-invariant
\begin{eqnarray}
\mbox{Tr}P'_{\alpha\beta}=\mbox{Tr}P_{\alpha\beta}.
\end{eqnarray}
We can combine the exponentials in Eq.(2.5) by using the identity $e^Ae^B=e^{A+B+1/2[A,B]+\cdot\cdot\cdot}$ repeatedly and obtain
\begin{eqnarray}
P_{\delta x,\delta y}=e^{i\delta x_{\mu}\delta y_{\nu}F^{\mu\nu}}
\end{eqnarray}
where $F_{\mu\nu}$ is the generally-covariant field-strength tensor $F_{\mu\nu}=\nabla_{\mu}A_{\nu}-i\nabla_{\nu}A_{\mu}-i\lambda [A_{\mu},A_{\nu}]$.\\
The Wilson action can now be formed from the plaquette operators just as in the flat spacetime case.\\
For simplicity, consider first the plaquette form by four vectors $\delta x^{\mu}=\langle 0,\delta x^1,0,0\rangle$ and $\delta y^{\nu}=\langle 0,0,\delta y^2,0\rangle$.

Consider now the expression
\begin{eqnarray}
\mbox{Tr}[P+P^{\dagger}]=-a^4\mbox{Tr}[F^{12}F_{12}] + \mbox{constant}
\end{eqnarray}
where $a$ is the \textit{physical} distance $a=\sqrt{g_{11}}\delta x^1=\sqrt{g_{22}}\delta y^2$ which we take to be constant along any side of a plaquette ($\delta x$ and $\delta y$ are \textit{coordinate} distance).\\
Note that the right-hand side of the equation above has a similar form with that of the kinetic term of the gauge field $F^{\mu\nu}F_{\mu\nu}$. To get this full kinetic term, sum Eq.(2.10) over all possible plaquettes
\begin{eqnarray}
\sum_{\square}=\sum_i\sum_{directions}
\end{eqnarray}
where we split the sum over all plaquettes into two parts: for each lattice site, sum over all directions, then sum over all lattice sites. The former produces the summation of the indices of the field-strength tensor $F^{\mu\nu}F_{\mu\nu}$. In the limit $a\rightarrow 0$, the summation over the lattice sites yields
\begin{eqnarray}
\lim_{a\rightarrow 0}\sum_ia^4=\int\mbox{d}^4x\sqrt{g}
\end{eqnarray}
where $g=$det$g_{\mu\nu}$.
Hence in this limit we have
\begin{eqnarray}
\lim_{a\rightarrow 0}\sum_{\square}\mbox{Tr}[P+P^{\dagger}]=-\lambda^2\int\mbox{d}^4x\sqrt{g}F^{\mu\nu}F_{\mu\nu}.
\end{eqnarray}
The right-hand side of the above equation is the covariant form of the kinetic term of the gauge field appropriate for curved spacetimes. Hence we can take the left-hand side expression as the appropriate Wilson action for curved spacetime.\\
Since $P^{\dagger}$ simply reverses the direction the plaquette is traversed, we can write the Wilson action as
\begin{eqnarray}
S=-\frac{1}{2\lambda^2}\sum_{\square}\mbox{Tr}P
\end{eqnarray}
where the sum includes both orientations of the plaquettes.\\
Eq.(2.14) can be useful for numerical simulations of lattice QCD in a curved spacetime. However, even without doing such numerical calculation, it is possible to extract an interesting result from the Wilson action above: it can be shown that for \textit{any} metric, a pair of gluons are always color-confined. This will be shown in the next section.

\subsection{Gluon color confinement}
From the definition of the link variable Eq.(2.1) or (2.2), it is not difficult to show that the relevant integral formulas $\int dU U_{ij}=0$, $\int dU U_{ij}U_{kl}=0$, $\int dU U_{ij}U^*_{kl}=\frac{1}{N}\delta_{ik}\delta_{jl}$ etc. remain true even in this curved spacetime formalism.\\
Since the usual integral formulas hold, we can the argue the following: for a Wilson loop $W_C$ of dimension $T\times L$ ($T$ is the wick-rotated time), the non-vanishing contribution to the path integral
\begin{eqnarray}
\int \mathcal{D}U W_C e^{-S},
\end{eqnarray}
expanded in the strong-coupling limit in powers of $\frac{1}{\lambda^2}$, is the configuration where the Wilson loop's interior is filled with plaquettes.\\
Since each plaquette contributes a factor $\frac{1}{\lambda^2}$ and there are $LT/a^2$ plaquettes inside the loop, we have
\begin{eqnarray}
\int \mathcal{D}U W_C e^{-S}\approx \bigg(\frac{1}{\lambda}\bigg)^{\frac{LT}{a^2}}.
\end{eqnarray}
This result shows a potential that is linear with separation $L$ and hence implies a color-confined glue-ball. Note that this result is true for a general metric including blackholes.\\
Also, the result (3.2) is consistent with the Cornell potential \cite{Eichten} except for the absence of the Yukawa potential which describes the color-Coulomb interaction of quarks. This is because we ignored the fermionic quarks in our calculation.\\
This result is also consistent with the 't Hooft relation with $\rho_{str}\equiv \frac{1}{a^2}$ as the coefficient of the dielectric term\cite{tHooft}. The presence of the scale $a$ means that the confinement phenomenon breaks scale invariance just as in the flat case\cite{Gaete06}.\\

\section{Quark in a Lattice}
\subsection{Quark Action}
In Section 2, we have written down the appropriate Wilson action for gluons in a curved spacetime. To complete the theoretical foundation of lattice gauge theory in a curved background, one also needs to write an appropriate lattice action for the quarks (Wilson fermions).

In curved spacetime, the Dirac equation is written as\cite{Birell82}
\begin{eqnarray}
i\gamma^{\alpha}\nabla_{\alpha}\Psi-m_Q\Psi = 0.
\end{eqnarray}
From here on, the first few Greek letters ($\alpha, \beta $...) denote vector indices in a local inertial frame while the middle letters ($\mu, \nu$...) denote vector indices in a general coordinate frame. This means that $\alpha$ and $\beta$ will be raised and lowered by the Minkowski metric $\eta_{\alpha\beta}$ while $\mu$ and $\nu$ by the general metric $g_{\mu\nu}$. The matrices $\{\gamma^{\alpha}\}$ in Eq.(4.1) are therefore the usual Dirac matrices in flat spacetime. The covariant derivative $\nabla_{\alpha}$ acting on the Dirac spinor $\Psi(x)$ is given by $\nabla_{\alpha}\equiv V^{\mu}_{\alpha}(\partial_{\mu}+\Gamma_{\mu})$ where the spin connection is given by $\Gamma\equiv \frac{1}{2}\Sigma^{\alpha\beta}V^{\nu}_{\alpha}(x)\partial_{\mu}V_{\beta\nu}(x)$, $V^{\mu}_{\alpha}$ is the vierbien and $\Sigma_{\alpha\beta}\equiv\frac{1}{4}[\gamma_{\alpha},\gamma_{\beta}]$ are the generators of Lorentz transformations in spinor representation.

As is usual, we will perform a Wick rotation before discretization: $t\rightarrow it$ and $\vec{\gamma}\rightarrow i\vec{\gamma}$. This changes the Dirac equation into
\begin{eqnarray}
g^{\mu\nu}\gamma_{\alpha}V^{\alpha}_{\mu}(\partial_{\nu}+\Gamma_{\nu})\Psi + m_Q\Psi=0.
\end{eqnarray}
The generators change to $\Sigma_{ij}=-\frac{1}{4}[\gamma_i,\gamma_j]$, $\Sigma_{0j}=i\frac{1}{4}[\gamma_0,\gamma_j]$ while the vierbiens are unchanged.

The discretization proceeds as usual: replace the derivatives by finite differences. The Dirac equation in matrix form becomes
\begin{eqnarray}
\sum_m \bigg\{&g^{\mu\nu}_n&\gamma_{\alpha}V^{\alpha}_{\mu n}\bigg[\frac{1}{2a}(\delta_{m,n+\nu}-\delta_{m,n-\nu})+\Gamma_{\nu n}\delta_{m,n}\bigg]+m_Q\delta_{m,n}\bigg\}\Psi_m=0
\end{eqnarray}
where in this notation $g^{\mu\nu}_n\equiv g^{\mu\nu}(x_n)$ and same goes for the spin connection and the vierbien.

If we include the coupling of the quarks with the gluons, Eq.(4.3) yields the propagator
\begin{eqnarray}
G^{-1}_{mn}(U,A)&=&\frac{1}{2a}g^{\mu\nu}_n\gamma_{\alpha}V^{\alpha}_{\mu n}\bigg[U_{\nu}(x_n)\delta_{m,n+\nu}-U^{\dagger}_{\nu}(x_m)\delta_{m,n-\nu}\bigg]\nonumber\\
&{}&{}+\big(g^{\mu\nu}_n\gamma_{\alpha}V^{\alpha}_{\mu n}\Gamma_{\nu n}+g^{\mu\nu}_n\gamma_{\alpha}V^{\alpha}_{\mu n}igA_{\nu n}+m_Q\big)\delta_{m,n}.
\end{eqnarray}
We see that the quark-gluon coupling terms are modified by the insertion of the factor $g^{\mu\nu}_nV^{\alpha}_{\mu n}$ while the mass term is modified into an \lq\lq effective mass\rq\rq  term $m_Q\rightarrow g^{\mu\nu}_n\gamma_{\alpha}V^{\alpha}_{\mu n}\Gamma_{\nu n}+m_Q$, each has a non-trivial dependence on $x_n$.

The path-integral for the Wilson fermion is then given by
\begin{eqnarray}
Z_{quark}\sim \int \prod_n\mbox{d}\Psi_n\prod_m\mbox{d}\bar{\Psi}_m\exp ^{-\bar{\Psi}_mG^{-1}_{mn}(U,A)\Psi_n}.
\end{eqnarray}
The fermion fields is usually integrated to yield the effective action
\begin{eqnarray}
S_{\mbox{eff}}[A]=S_0[A]-\mbox{Tr}\mbox{ln} G^{-1}(U,A)
\end{eqnarray}
where $S_0[A]$ is the action for gluons.\\
This equation can be used as a starting point for numerical simulations. The computation of the last term might be slow but there are methods to handle this.\cite{Fucito81, Scalapino81}

\subsection{The fermion-doubling Problem}
It is well-known that lattice gauge theories typically suffer from the fermion-doubling problem. This happens  when the propagator has an extra pole that falls within the Brillouin zone. For the propagator in Eq.(3.4), whether this pole exist depends in general on the geometry of the background spacetime $g_{\mu\nu}$. In this section, we consider the special case of Friedman-Robertson-Walker(FRW) metric to investigate the fermion-doubling problem for lattice QCD in curved spacetime.\\
For this case, it is more convenient to use the convention $(+---)$ for the spacetime signature so that the propertime element is given by d$s^2=$d$t^2-a^2(t)$(d$x^2+$d$y^2+$d$z^2$). Since the time variable $t$ appears in the function $a(t)$, it is more convenient to apply Wick rotation on the spatial part $\vec{x}\rightarrow i\vec{x}$. The propertime now has a Euclidean signature d$s^2=$d$t^2+a^2(t)$(d$x^2+$d$y^2+$d$z^2$). Accordingly, we must also rotate the gamma matrix $\gamma^{\hat{0}}\rightarrow -i\gamma^{\hat{0}}$ where a hat in the index zero means that it is a local frame index. The Dirac equation in curved spacetime then becomes 
\begin{eqnarray}
\gamma^{\alpha}V^{\mu}_{\alpha}(\partial_{\mu}+\Gamma_{\mu})\Psi-m\Psi =0.
\end{eqnarray}
Following the discretization done in Section 3.1, the propagator becomes
\begin{eqnarray}
(G^{-1})_{nm}=g^{\mu\nu}_n\gamma_{\alpha}V^{\alpha}_{\mu n}\bigg[\frac{1}{2a}(\delta_{m,n+\nu}-\delta_{m,n-\nu})+\Gamma_{\nu n}\delta_{n,m}\bigg]-m\delta_{n,m}.
\end{eqnarray}
The Fourier transform of the above is
\begin{eqnarray}
G^{-1}(p)&\equiv& \frac{1}{a^4V_4}\sum_{nm}(G^{-1})_{nm}e^{-ip\cdot (x_m-x_n)}\\
&=&i\sum_{\nu}p_{\nu}\frac{\sin (p_{\nu}a)}{p_{\nu}a}\frac{1}{V_4}\sum_n a^4\gamma^{\nu}_n+\frac{1}{V_4}\sum_na^4\gamma^{\nu}_n\Gamma_{\nu}-m
\end{eqnarray}
where $V_4\equiv \sum a^4$ is the volume of our (finite) lattice.

Note that getting the continuum limit $a\rightarrow 0$, Eq.(3.9) gives an alternative approach to obtain the momentum-space propagator for fermions in curved spacetime
\begin{eqnarray}
G^{-1}(p)=ip_{\nu}\tilde{\gamma^{\nu}}+M
\end{eqnarray}
where $\tilde{\gamma^{\nu}}\equiv V_4^{-1}\int\mbox{d}^4x\sqrt{g}\gamma^{\nu}(x)$ and $M\equiv V_4^{-1}\int\mbox{d}^4x\sqrt{g}\gamma^{\nu}(x)\Gamma_{\nu}(x)-m$.

Let us consider the specific orthonormal basis whose coordinate components are given by $e_{\hat{0}}^{\mu}=\langle 1,0,0,0\rangle$, $e_{\hat{1}}^{\mu}=\langle 0,a^{-1}(t),0,0\rangle$, $e_{\hat{2}}^{\mu}=\langle 0,0,a^{-1}(t),0\rangle$ and $e_{\hat{3}}^{\mu}=\langle 0,0,0,a^{-1}(t)\rangle$. The vierbein is then diag$(1,a^{-1},a^{-1},a^{-1})$. It can be shown that in this metric, all the spin connection components vanish $\Gamma_{\mu}\equiv \frac{1}{2}\Sigma^{\alpha\beta}V^{\nu}_{\alpha}\partial_{\mu}V_{\beta\nu}=0$. 

Let us now investigate the zeros of $G^{-1}(p)=0$ which we write as
\begin{eqnarray}
i\gamma_{\alpha}F^{\alpha}+m&=&0\\
F^2+m^2&=&0
\end{eqnarray}
where 
\begin{eqnarray}
F^{\alpha}\equiv \sum_{\nu}a^{-1}\sin (p_{\nu}a)V_4^{-1}\sum_na^4g^{\mu\nu}_nV^{\alpha}_{\mu n}-iV_4^{-1}\sum_na^4g^{\mu\nu}_nV^{\alpha}_{\mu n}\Gamma_{\nu n}.
\end{eqnarray}
To get Eq.(3.12) we multiplied Eq.(3.11) by $-i\gamma_{\beta}F^{\beta}+m$ and used the identity $\gamma_{\alpha}\gamma_{\beta}=\frac{1}{2}\{\gamma_{\alpha},\gamma_{\beta}\}=\eta_{\alpha \beta}$.

For the purpose of illustrating fermion-doubling, let us consider the massless case $m=0$. Eq.(3.12) then simplifies to
\begin{eqnarray}
\eta_{\alpha\beta}\sum_{\nu}\frac{1}{a}\sin (p_{\nu} a)B^{\nu \alpha}\sum_{\lambda}\frac{1}{a}\sin (p_{\lambda}a)B^{\lambda\beta}=0
\end{eqnarray}
where $B^{\alpha\mu}\equiv\mbox{diag}(1,b,b,b)$, $b\equiv V^{-1}_4\int\mbox{d}^4x\sqrt{g}a^{-1}(t)=V^{-1}_4\int\mbox{d}^3x\mbox{d}ta^2(t)$ and $V_4=\int\mbox{d}^4x\sqrt{g}=\int\mbox{d}^3x\mbox{d}ta^3(t)$. For a vacuum dominated universe at large times $a(t)\sim e^{Ht}$, where $H$ is the Hubble \lq\lq constant\rq\rq, and we have
\begin{eqnarray}
b=\lim_{T\rightarrow\infty}\frac{\int^T\mbox{d}ta^2(t)}{\int^T\mbox{d}ta^3(t)}\sim \lim_{T\rightarrow\infty}e^{-HT}=0.
\end{eqnarray}
Eq.(3.14) then reduces to 
\begin{eqnarray}
\sin^2(p_0a)=0
\end{eqnarray}
which has a solution $p_0=\frac{\pi}{a}$ at the boundary of the 4-dimensional Brilluoin zone. Note that if we interpret $p_0$ as energy, then this solution looks like a flat band inside the 3-dimensional $\vec{p}$-space Brilluoin zone.

This specific example shows that the fermion-doubling problem can occur in the lattice gauge theory in curved spacetime. Due to the non-trivial appearance of the factors like $g_{\mu\nu}$, $V^{\mu}_{\alpha}$ and $\Gamma_{\mu}$ however, the doubling of the fermions depends on the specific spacetime geometry under consideration.

\section{Final remarks}
In this paper, it was shown that a lattice gauge theory can be consistently formulated in curved spacetime background. For the gluons, we have derived the Wilson action which reduces to the appropriate covariant action in the continuum limit. Even without performing a numerical calculation, it was shown analytically that in the strong-coupling limit, two gluons propagating in a curved spacetime are always color-confined. This result does not use any effective low-energy model and is valid for any bakcground geometry including blakchole spacetime. Primordial blackholes therefore, do not radiate free gluons. 

The lattice action for the quarks was also derived. This was done by solving the discretized propagator. We have seen that in curved spacetime, there is an additional contribution to the mass matrix of the form $g^{\mu\nu}_n\gamma_{\alpha}V^{\alpha}_{\mu n}\Gamma_{\nu n}+m_Q$ and a new factor $g^{\mu\nu}_nV^{\alpha}_{\mu n}$ to the quark kinetic term. It was also shown in the specific case of an FRW metric that fermion-doubling can appear. 

This paper lays down the theoretical foundation for lattice QCD in curved spacetime. Future works can now be directed on numerical calculations. This is important, for example, in determining how gravity affects the potential between two quarks and how gravity affects confinement. To do a numerical calculation, one must commit to a specific metric, for example, within the neighborhood of the blackhole horizon or FRW model. We have shown that the fermion-doubling problem can appear. Hence one must be careful to eliminate this extra fermion, for example by adding a Wilson parameter to make this spurious fermion infinitely massive or by using staggered fermions\cite{Banks, Susskind}. One can then use the standard numerical techniques like Monte Carlo and Metropolis algorithms\cite{Creutz, Rebbi}.


\begin{thebibliography}{0}
\bibitem{Pines}
eds. D. Pines, R. Tamagaki and S. Tsuruta, {\it Conference proceedings: the structure and evolution of neutron stars},
(Addison-Wesley publishing company, 1992).

\bibitem{Alford}
M. G. Alford, A. Schmitt, K. Rajagopal and T. Schafer, {\it Rev. Mod. Phys.} {\bf 80}, 1455-1514 (2008).

\bibitem{Gass}
R. Gass, {\it Phys. Rev. D} {\bf 27}, 2893-2905 (1983).

\bibitem{Davies81}
P. C. W. Davies, edited by R. Penrose, C. J. Isham and D. Sciama, {\it Quantum gravity 2: a second Oxford symposium},
(Oxford University Press, 1981).

\bibitem{Hartle}
J. B. Hartle and S. W. Hawking, {\it Phys. Rev. D} {\bf 13}, 2188-2203 (1976).

\bibitem{Flachi13}
A. Flachi, {\it Phys. Rev. D} {\bf 88}, 041501 (2013).

\bibitem{Gaete}
P. Gaete and E. Spallucci, {\it Phys. Rev. D} {\bf 77}, 027702 (2008).

\bibitem{Gaete06}
P. Gaete and E. Guendelman, {\it Phys. Lett. B} {\bf 640}, 201-204 (2006).

\bibitem{Gaete07}
P. Gaete, E. Guendelman and E. Spalluci, {\it Phys. Lett. B} {\bf 649}, 218-224 (2007).

\bibitem{Vasihoun14}
M. Vasihoun, E. Guendelman, {\it arXiv:1406.0203v1} {\bf [gr-qc]}, (2014).

\bibitem{Eichten}
E. Eichten, K. Gottfried, T. Kinoshita, K. D. Lane and T. M. Yan, {\it Phys. Rev. D} {\bf 17}, 3090-3117 (1978).

\bibitem{Kharzeev04}
D. Kharzeev, E. Levin and K. Tuchin, {\it Phys. Rev. D} {\bf 70}, 054005 (2004).

\bibitem{Migdal}
A. A. Migdal and M. A. Shifman, {\it Phys. Lett. B} {\bf 114}, 445-449 (1982).

\bibitem{Sasagawa}
S. Sasagawa and H. Tanaka, {\it Prog. Theor. Phys.} {\bf 128}, 925-939 (2012).

\bibitem{tHooft}
G. 'tHooft, {\it Nucl. Phys. B, Proc. Suppl.} {\bf 121}, 333-340 (2003).

\bibitem{Kharzeev}
D. Kharzeev, E. Levin and K. Tuchin, {\it Phys. Lett. B} {\bf 547}, 21-30 (2002).

\bibitem{Gaete01}
P. Gaete, {\it Phys. Lett. B} {\bf 515}, 382-386 (2001).

\bibitem{Birell82}
N. D. Birell and  P. C. W. Davies, in {\it Quantum Fields in Curved Space},
(Cambridge University Press, 1982).

\bibitem{Fucito81}
F. Fucito, E. Marinari, G. Parisi, C. Rebbi, {\it Nuc. Phys. B} {\bf 180}, 369-377 (1981).

\bibitem{Scalapino81}
D. J. Scalapino, R. L. Sugar, {\it Phys. Rev. Lett.} {\bf 46}, 519-521 (1981).

\bibitem{Banks}
T. Banks, S. Raby, L. Susskind, J. Kogut, D. R. T. Jones, P. N. Scharbach and D. K. Sinclair, {\it Phys. Rev. D} {\bf 15}, 1111-1127 (1977).

\bibitem{Susskind}
L. Susskind,  {\it Phys. Rev. D} {\bf 16}, 3031-3039 (1977).


\bibitem{Creutz}
M. Creutz, L. Jacobs and C. Rebbi, {\it Phys. Rep.} {\bf 95}, 201-282 (1983).


\bibitem{Rebbi}
C. Rebbi, {\it Phys. Rev. D} {\bf 21}, 3350-3359 (1980).

\end{thebibliography}
\end{document}